%% file: main.tex
\documentclass[conference]{IEEEtran}
\IEEEoverridecommandlockouts
\usepackage{cite}
\usepackage{amsmath,amssymb,amsfonts}
\usepackage{graphicx}
\usepackage{textcomp}
\usepackage{xcolor}
\usepackage{xspace}
\usepackage{algorithm}
\usepackage{algpseudocode}
\algnewcommand\algorithmicinput{\textbf{Input:}}
\algnewcommand\algorithmicoutput{\textbf{Output:}}
\algnewcommand\Input{\item[\algorithmicinput]}
\algnewcommand\Output{\item[\algorithmicoutput]}
\usepackage{tabularx}
\usepackage{svg}
\usepackage{subcaption}
\usepackage{enumitem}
\usepackage{multirow}
\usepackage{booktabs}
\usepackage{array}
\usepackage{stfloats}
\usepackage{float}
\usepackage[section]{placeins}
\usepackage{amsthm}
\usepackage{adjustbox}
\usepackage{url}
\usepackage{comment}
\usepackage[T1]{fontenc}
\usepackage{soul}
\sethlcolor{yellow}
\def\BibTeX{{\rm B\kern-.05em{\sc i\kern-.025em b}\kern-.08em
    T\kern-.1667em\lower.7ex\hbox{E}\kern-.125emX}}

\newcommand{\cmark}[1]{\textcircled{#1}}

\newcommand{\zn}[1]{}
\newcommand{\tk}[1]{}
\newcommand{\note}[1]{}
\newcommand{\info}[1]{}

\newcommand{\name}{GreenDyGNN\xspace}

\begin{document}
\title{\name: Runtime-Adaptive Energy-Efficient Communication for Distributed GNN Training}

\author{
\IEEEauthorblockN{Arefin Niam}
\IEEEauthorblockA{
Tennessee Technological University\\
Cookeville, Tennessee, USA\\
aniam42@tntech.edu
}
\and
\IEEEauthorblockN{Tevfik Kosar}
\IEEEauthorblockA{
University at Buffalo\\
Buffalo, New York, USA\\
tkosar@buffalo.edu
}
\and
\IEEEauthorblockN{M. S. Q. Zulkar Nine}
\IEEEauthorblockA{
Tennessee Technological University\\
Cookeville, Tennessee, USA\\
mnine@tntech.edu
}
}

\maketitle
\begin{abstract}
Distributed GNN training is dominated by remote feature fetching, which can be very costly. Multi-hop neighborhood sampling crosses partition boundaries and triggers fine-grained RPCs whose fixed initiation cost and GPU-stall latency waste energy. Prior systems try to reduce this overhead with presampling and static caching, but cache policies cannot react to runtime network variation. We show that under time-varying congestion, static caching can increase energy by up to 45\% because a fixed rebuild schedule is insufficient. We present \name, which formulates cache window management as a sequential decision problem. \name performs intra-epoch cache rebuilds and uses a Double-DQN agent, trained in a calibrated simulator with domain-randomized congestion, to adapt rebuild window size and per-owner cache allocation at each boundary. An asynchronous double-buffered pipeline makes adaptation effectively free. Under congestion, \name cuts total energy by up to 43\% over Default DGL and 4--24\% over the best static policy, while closely matching the optimum under clean conditions.
\end{abstract}

\begin{IEEEkeywords}
Distributed GNN training, energy efficiency, runtime adaptation, reinforcement learning, communication optimization.
\end{IEEEkeywords}

\input{section-introduction.tex}
\input{section-motivation}
\input{section-relatedwork.tex}
\input{section-proposedwork.tex}
\input{section-experimentation.tex}

\section{Conclusion}

We presented \name, a runtime-adaptive system for energy-efficient distributed GNN training. The core insight is that the cache window size is a control variable that must be adapted online as network conditions evolve, not a hyperparameter to be tuned offline. \name formalizes this as a sequential decision problem and solves it with a Double-DQN agent trained via sim-to-real transfer under domain-randomized congestion. An asynchronous double-buffered pipeline makes adaptation overhead-free. Under realistic time-varying congestion the learned policy consistently outperforms every static configuration, absorbing up to 26 percentage points of congestion overhead on the largest graph tested. Under clean conditions it converges to the static optimum.

Future work includes extending the formulation to heterogeneous hardware (mixed GPUs, RDMA), adding GPU frequency scaling as an action dimension, and online fine-tuning of the simulator from production telemetry.

\section*{Acknowledgments}

An AI writing assistant (Claude, Anthropic)~\cite{claude2024} was used to improve grammar and readability across all sections of this paper. All technical content, experimental design, implementation, and results were produced entirely by the authors, who verified the final text for accuracy.

\bibliographystyle{IEEEtran}
\bibliography{references}

\end{document}

%% file: section-introduction.tex
\section{Introduction}
\label{sec:intro}

Graph neural networks (GNNs) have become a standard framework for learning on relational data, with applications in molecular discovery, recommendation, fraud detection, and social network analysis~\cite{kipf2017semi, hamilton2017inductive}. As real-world graphs grow to billions of edges, training moves to distributed clusters that partition the graph across workers and synchronize model updates through data-parallel execution~\cite{zheng2020distdgl, gandhi2021p3, jia2020roc, zheng2022bytegnn}. In this setting, each mini-batch expands a multi-hop neighborhood that often crosses partition boundaries, forcing workers to fetch remote node features through fine-grained RPCs. These fetches consume a large fraction of per-step time on commodity clusters~\cite{niam2025rapidgnn, sarkar2024massivegnn} and waste energy in a particularly costly way: while the worker waits for remote features, the GPU continues to draw near-idle power but cannot make progress. This makes distributed GNN training fundamentally different from conventional distributed DNN training, where communication is dominated by gradient synchronization. Here, the main bottleneck is irregular, data-dependent feature movement.

A growing body of work reduces this overhead through caching, prefetching, and communication consolidation~\cite{lin2020pagraph, kaler2023salient, liu2023bgl, niam2025rapidgnn}. Among these, RapidGNN~\cite{niam2025rapidgnn} shows that presampling the training trace once per epoch and caching frequently accessed remote features can substantially reduce communication cost and energy. However, this cache policy remains static within the epoch. It fixes both rebuild timing and cache composition in advance, which means it cannot react when runtime network conditions shift.

That rigidity becomes costly under congestion. On a 4-node cluster, injecting just 4\,ms of additional latency on a single inter-node link shifts the energy-optimal cache rebuild regime and increases per-epoch energy by more than 60\% under the best fixed policy. This kind of perturbation is not unusual in shared clusters, where co-tenant jobs, cross-rack traffic, and transient switch contention introduce millisecond-scale variation over time. Once conditions change, a static cache policy continues executing a communication plan shaped for an earlier network state: rebuild timing is no longer well matched to actual fetch cost, cache space is not directed where it is most valuable, and remote stalls leave the GPU waiting. The challenge becomes even harder when congestion is time-varying or affects multiple links at once, because the best rebuild interval and cache allocation then depend on a multi-dimensional congestion state that changes over seconds or minutes. No single fixed policy can track this reliably.

This paper presents \textbf{\name}, which turns cache window management into an online control problem. Unlike prior systems that rely on static epoch-level caching, \name introduces fine-grained window-based cache rebuilds within an epoch and adapts them at runtime. At each rebuild boundary, \name observes per-owner fetch latencies and cache statistics, then selects both the rebuild window size $W$ and the per-owner cache allocation for the next window. Because each decision affects many future training steps under non-stationary conditions, and because window size and allocation bias interact in ways that simple heuristics cannot fully capture, this naturally becomes a sequential decision problem with delayed and coupled effects. \name addresses it with a lightweight Double-DQN agent trained in a calibrated simulator under domain-randomized congestion and deployed into the live training loop with negligible overhead per decision. To avoid costly on-cluster RL training, we use a sim-to-real design: short profiling runs calibrate the simulator, and domain randomization exposes the policy to congestion patterns beyond those seen in calibration. An asynchronous double-buffered prefetching pipeline ensures that cache rebuilds, at whatever frequency the policy selects, do not stall the GPU.

\name makes the following contributions:
\begin{enumerate}
\item It identifies \textbf{runtime network variability} as a first-class systems problem for energy-efficient distributed GNN training, showing that even modest link-level congestion can shift the energy-optimal operating point and substantially increase energy consumption under fixed cache policies (Section~\ref{sec:motivation}).

\item It introduces \textbf{fine-grained window-based cache rebuilding} for distributed GNN training and formulates runtime cache adaptation as a \textbf{Markov decision process}, solved with a Double-DQN agent trained through \textbf{sim-to-real transfer} with domain randomization over congestion patterns, temporal behavior, and measurement noise (Section~\ref{sec:design}).

\item It designs an \textbf{asynchronous double-buffered pipeline} that decouples cache rebuilds from the training loop, making runtime adaptation effectively overhead-free at the decision timescale (Section~\ref{sec:impl}).

\item Under realistic time-varying congestion on a 4-node cluster, \name reduces total energy by up to 43\% relative to Default DGL and outperforms the strongest static caching baseline, RapidGNN, by 4--24\% across all 9 configurations. On the largest graph, the learned policy absorbs up to 26 percentage points of congestion overhead that static caching cannot. Under clean conditions (i.e., no congestion), \name matches the best fixed operating point.
\end{enumerate}

%% file: section-motivation.tex
\section{Motivation}
\label{sec:motivation}

We profile a 4-node Chameleon Cloud~\cite{keahey2019chameleon} cluster running DistDGL GraphSAGE~\cite{hamilton2017inductive, zheng2020distdgl} on OGBN-Products (METIS-partitioned, PyTorch DDP~\cite{paszke2019pytorch}) and identify two physical mechanisms that make remote feature fetching the dominant energy consumer. Together, they explain why \emph{static} cache policies are brittle under runtime network variability and why distributed GNN training needs adaptive cache control.

\subsection{Per-RPC Initiation Overhead}
\label{subsec:init_overhead}

Every remote feature fetch pays a \emph{fixed initiation cost} $E_{\text{init}}$, which includes CPU interrupt handling, kernel crossings, socket-buffer allocation, and TCP/IP protocol processing. RDMA bypasses the kernel and removes most software overhead, but each transfer still requires posting a work request, signaling the NIC, and waiting for a completion event. These steps take on the order of 1--2~$\mu$s regardless of how many bytes are sent, so for the tens to low hundreds of remote nodes that GNN sampling typically requests per mini-batch, the fixed per-transfer cost still outweighs the payload cost. On top of this, each fetch pays a \emph{variable payload cost} $E_{\text{payload}} \propto N \cdot F_b$, where $N$ is the number of remote nodes and $F_b$ is the per-node feature size in bytes. For the small, irregular transfers produced by GNN neighborhood sampling, the fixed component dominates.

\begin{figure}[t]
  \centering
  \includegraphics[width=\columnwidth]{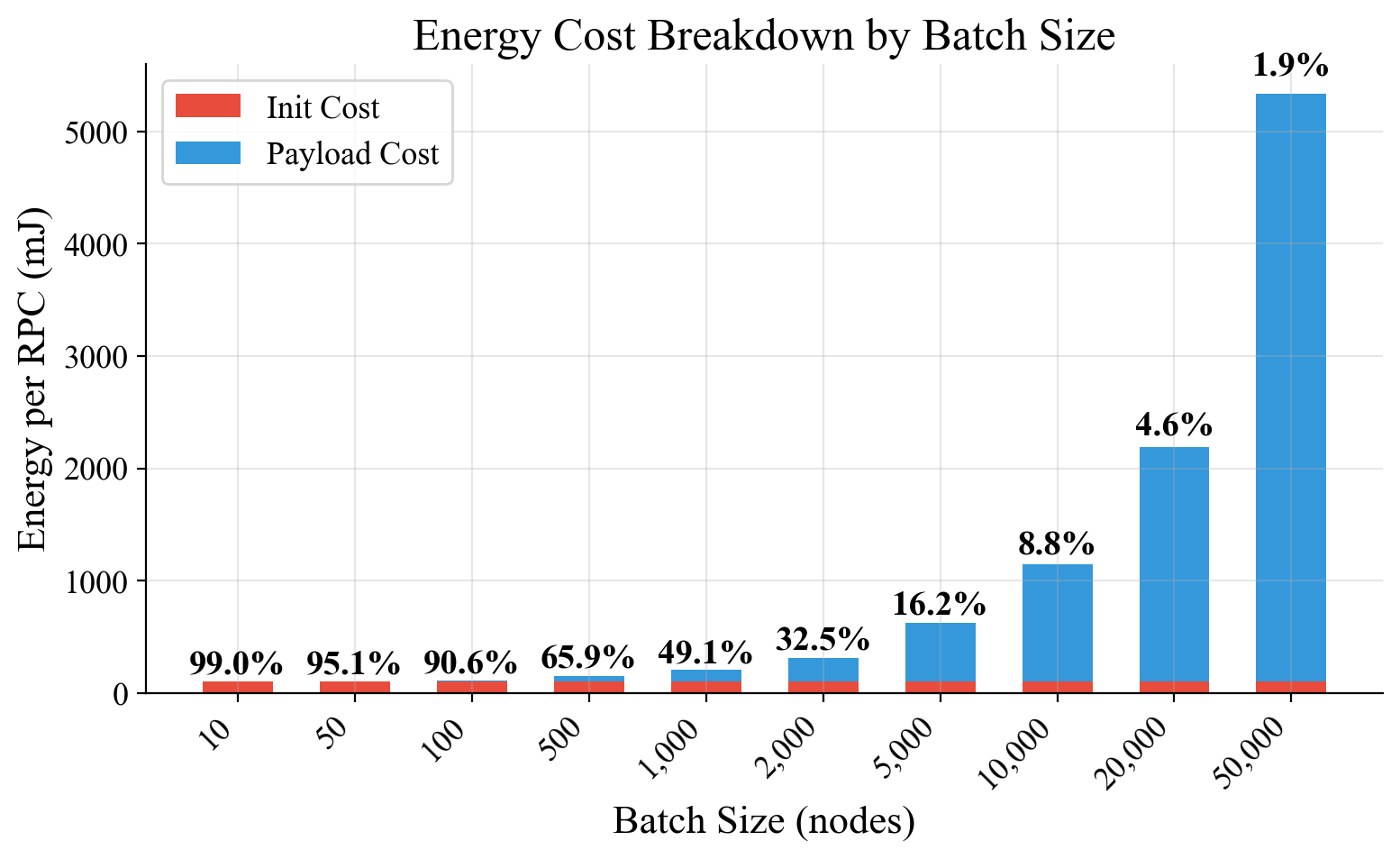}
  \caption{Per-RPC energy decomposed into initiation (red) and payload (blue) cost as a function of batch size. Percentages indicate the initiation share. At GNN-typical request sizes (tens to hundreds of nodes), initiation accounts for 90--99\% of total energy. Payload cost does not dominate until batch sizes exceed ${\approx}$1{,}000 nodes, well above what distributed GNN training produces per mini-batch.}
  \label{fig:crossover}
\end{figure}

We quantify this by profiling DistTensor RPCs across a range of payload sizes on our 25\,Gbps cluster. Figure~\ref{fig:crossover} breaks down per-RPC energy cost into initiation and payload components as batch size grows from 10 to 50{,}000 nodes. At GNN-typical request sizes, initiation accounts for 90--99\% of total per-RPC energy. Even at 1{,}000 nodes, initiation still consumes roughly half the energy. The crossover at which payload cost begins to dominate does not occur until batch sizes exceed approximately 1{,}000 nodes, far above the per-batch remote set size observed in practice. Distributed GNN training therefore operates largely in the initiation-dominated regime, paying a high fixed energy cost to move relatively little data.

The immediate implication is that many small remote fetches should be consolidated into fewer, larger transfers. This motivates rebuilding a cache over an upcoming window of batches: if the \textit{hot} remote nodes needed in the next $W$ steps are fetched together in one bulk transfer, the system can amortize the fixed initiation cost across many future accesses. The larger the window, the stronger this amortization becomes, although, as we show next, larger windows also make the cache less fresh.

\subsection{GPU Stall Energy}
\label{subsec:stall_energy}

Modern datacenter GPUs draw 40--80\,W even when idle, a significant fraction of their 250--400\,W TDP (Thermal Design Power). In distributed GNN training, the GPU frequently stalls while the CPU thread resolves remote features. The forward pass cannot begin until all input-node features are available, and the last remote RPC to return determines the stall duration, creating a straggler effect across remote owners. Every millisecond of stall time therefore translates directly into wasted energy at the GPU's idle power draw.

Pipelining and prefetching can overlap communication with computation and improve utilization under normal conditions. \name's asynchronous pipeline (Section~\ref{sec:pipeline}) removes most GPU idle time when the network has no contention. However, overlap alone does not reduce the \emph{total volume} of remote data fetched or the \emph{total energy} consumed by those transfers. When congestion inflates RPC latencies, the prefetcher can no longer resolve future batches quickly enough, and stalls reappear. Reducing stall energy under congestion requires reducing expensive remote misses themselves, either by refreshing the cache more frequently or by steering more cache capacity toward the congested owner. Both responses require runtime awareness of network conditions.

\subsection{Why Static Policies Fail}
\label{subsec:why_adaptive}

The two mechanisms above create a fundamental tradeoff. Larger cache-rebuild intervals amortize initiation costs over more steps, reducing rebuild overhead per batch, but they also make cached contents less fresh, which increases miss rate, remote fetch cost, and GPU stall time. Smaller rebuild intervals improve cache freshness and reduce misses, but they pay rebuild overhead more often. Any static cache policy therefore fixes a single operating point in this tradeoff.

RapidGNN-style epoch-level caching fixes that operating point once per epoch. To expose the underlying tension more clearly, consider a finer-grained window-based rebuild regime in which the cache is rebuilt every $W$ steps. Under clean network conditions, the energy-optimal window balances amortized rebuild cost against miss cost at a stable point. Under congestion, however, the per-miss cost $t_{\text{miss}}$ rises, the miss-energy term grows disproportionately, and the optimal operating point shifts toward smaller windows that tolerate more frequent rebuilds in exchange for fewer expensive remote fetches.

On OGBN-Products, the energy-optimal window shifts from $W^*{=}16$ under clean conditions to $W^*{\approx}8$ under 4\,ms single-link congestion. Operating at the wrong window inflates energy by over 60\%. Conversely, operating at $W{=}8$ under clean conditions wastes energy on unnecessary rebuilds. No single fixed choice is correct across both regimes.

The challenge becomes harder when congestion is time-varying or affects multiple links simultaneously. In that setting, the best rebuild interval and cache allocation become functions of a multi-dimensional congestion vector $\boldsymbol{\sigma} = (\sigma_1, \ldots, \sigma_{P-1})$ that changes over seconds to minutes. A threshold rule can approximate the right response under simple single-link congestion, but it degrades quickly once patterns become multi-link, asymmetric, or time-varying. What is needed is a policy that maps the current network state to the best joint decision over rebuild window and cache allocation. That is the adaptive control problem \name is designed to solve.

%% file: section-relatedwork.tex
\section{Related Work}
\label{sec:related}

\textbf{Distributed GNN systems.}
DistDGL~\cite{zheng2020distdgl} introduced the distributed graph store and DistTensor RPC abstraction that enable training on billion-scale partitioned graphs, forming the substrate on which \name builds. P3~\cite{gandhi2021p3} overlaps inter-partition communication with GPU computation in a layer-wise pipeline. Roc~\cite{jia2020roc} dynamically repartitions to balance computation. ByteGNN~\cite{zheng2022bytegnn} combines high sampling parallelism with workload-aware partitioning for large speedups over DistDGL. BNS-GCN~\cite{wan2022bns} reduces boundary communication in full-graph training by randomly sampling boundary nodes. GNNLab~\cite{yang2022gnnlab} decouples sampling from training across CPU and GPU and uses pre-sampling to reduce mini-batch generation variance. Across these systems, the main goals are throughput and scalability. None explicitly targets energy-efficient runtime adaptation to changing network conditions.

\textbf{Feature caching and prefetching.}
PaGraph~\cite{lin2020pagraph} caches frequently accessed features in GPU memory using offline frequency analysis. DUCATI~\cite{zhang2023ducati} extends this idea with a dual-cache architecture and counting-Bloom-filter-based frequency estimation. BGL~\cite{liu2023bgl} builds a multi-tier cache hierarchy spanning GPU, CPU, and SSD with graph-aware I/O scheduling. SALIENT++~\cite{kaler2023salient} provides theoretical analysis of miss rates under static caching. MassiveGNN~\cite{sarkar2024massivegnn} introduces asynchronous prefetching for massively connected graphs. RapidGNN~\cite{niam2025rapidgnn} shows that epoch-level presampling and static hot-feature caching can substantially reduce communication and energy in distributed GNN training. Orthogonally, DGCL~\cite{cai2021dgcl} fuses small messages and balances load across heterogeneous interconnects to reduce training time. The common limitation is that these systems rely on fixed cache or communication policies decided offline or at deployment time. In contrast, \name introduces intra-epoch window-based cache rebuilding and adapts both the rebuild window and cache allocation online using a learned policy targeted at energy.

\textbf{Energy-efficient ML training.}
The environmental cost of large-model training has drawn substantial attention~\cite{strubell2019energy, schwartz2020green, patterson2022carbon}. At the systems level, Zeus~\cite{you2023zeus} jointly optimizes GPU power limits and batch size for single-GPU DNN training, achieving large energy reductions. Perseus~\cite{chung2024perseus} studies energy bloat from pipeline bubbles in large-model training and exposes an iteration-time versus energy Pareto frontier. EnvPipe~\cite{choi2023envpipe} modulates GPU frequency during non-critical pipeline stages. These systems target compute-side energy in DNN training. In distributed GNN training, however, a major source of waste is network-induced GPU idle power during remote feature stalls. RapidGNN~\cite{niam2025rapidgnn} is the closest prior work in this space because it explicitly measures and reduces energy in distributed GNN training. \name builds on that direction, however, replaces static epoch-level caching with an adaptive online policy.

\textbf{RL for systems and sim-to-real transfer.}
Pensieve~\cite{mao2017pensieve} for adaptive bitrate streaming, Decima~\cite{mao2019decima} for cluster scheduling, and learned device placement~\cite{mirhoseini2017device} show that RL can outperform hand-tuned heuristics in dynamic systems settings. \name applies this idea to cache control in distributed GNN training, where decisions are infrequent, their effects persist across many future steps, and energy is the primary optimization target. To make training practical, \name uses sim-to-real transfer through domain randomization~\cite{tobin2017domain}. This idea was first developed in robotics and later demonstrated at scale by systems such as OpenAI's Dactyl~\cite{openai2020dactyl}. In our setting, domain randomization exposes the policy to diverse congestion patterns during simulator training so that it generalizes to unseen runtime conditions on the real cluster. Graph partitioning methods such as METIS~\cite{karypis1998fast} are orthogonal to our contribution: \name operates after partitioning and reduces the energy cost of the cross-partition communication that still remains.

%% file: section-proposedwork.tex
\section{\name Design}
\label{sec:design}

\begin{figure}[t]
  \centering
  \includegraphics[width=0.98\columnwidth]{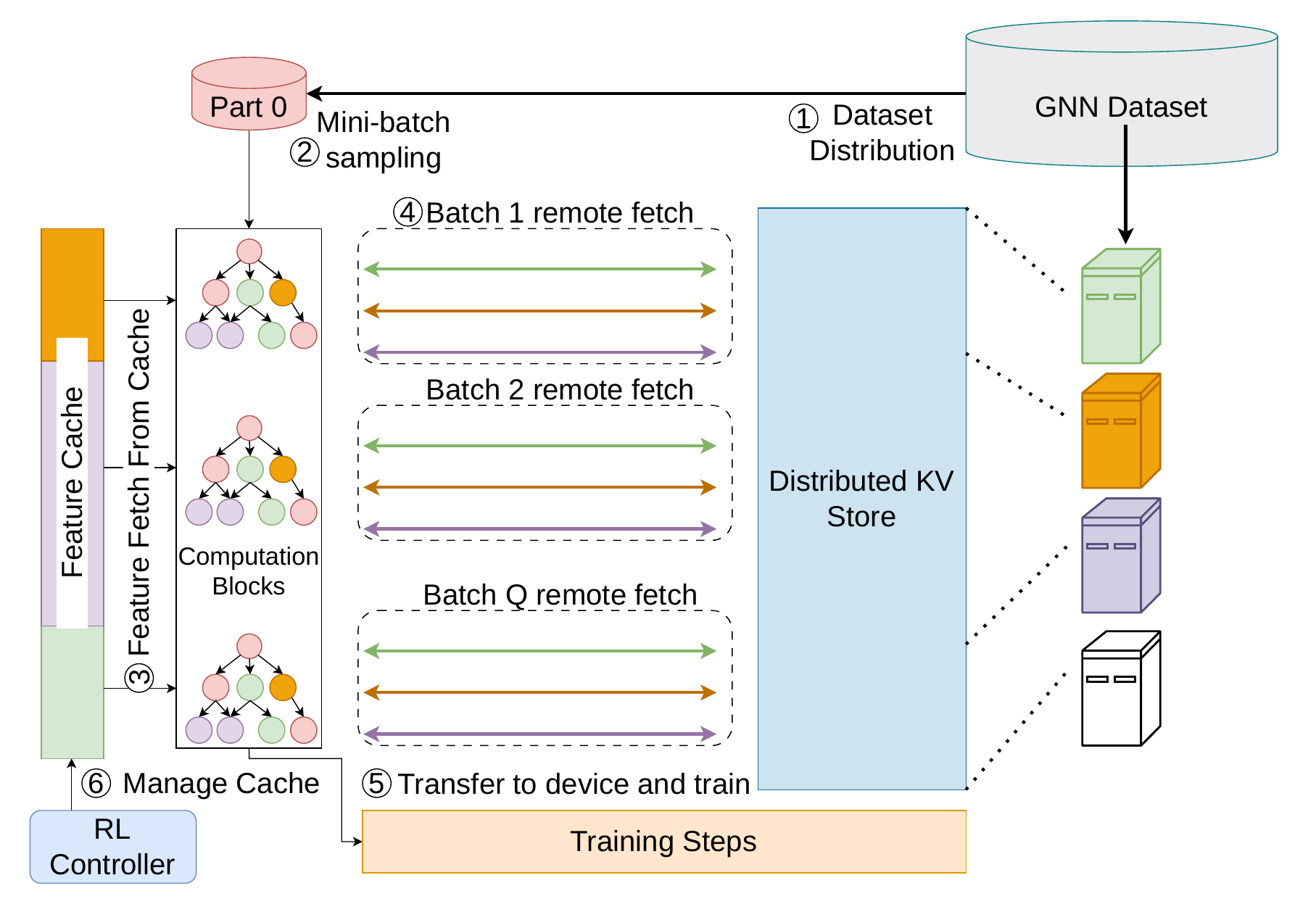}
  \caption{\name runtime training pipeline (single-worker view).
  \cmark{1}~The graph dataset is partitioned and distributed across nodes.
  \cmark{2}~A background thread samples mini-batches from the local partition.
  \cmark{3}~Sampled computation blocks retrieve node features from the local
  feature cache. \cmark{4}~Cache misses trigger batched remote fetches from the
  distributed KV store, with up to $Q$ batches resolved concurrently across
  owners. \cmark{5}~Resolved features are transferred to the GPU for training.
  \cmark{6}~The RL controller observes fetch latencies and cache statistics,
  then selects the next rebuild window $W$ and per-owner cache allocation at each
  rebuild boundary.}
  \label{fig:arch}
\end{figure}

\name operates in three phases. First, an offline calibration phase,
run once per cluster, builds a lightweight cost model and simulator from
measured communication and power data. Second, a Double-DQN agent is
trained in that simulator under domain-randomized congestion profiles,
using the state inputs and action outputs shown in
Figure~\ref{fig:rl-agent}. Third, the trained policy is deployed inside
the asynchronous prefetching pipeline shown in Figure~\ref{fig:arch},
where it selects the rebuild window and per-owner cache allocation at
each cache rebuild boundary with no measurable overhead on the critical
path.

\subsection{Cost Model}
\label{sec:model}

Consider a distributed GNN trainer with $P$ partitions, batch size $B$,
and rebuild window $W$. Prior systems such as RapidGNN rebuild the
remote-feature cache once per epoch using epoch-level presampling. In
contrast, \name introduces fine-grained window-based rebuilding within
an epoch. Every $W$ steps, a background thread rebuilds the cache using
the hot remote nodes expected in the upcoming window. The wall-clock
time of a single training step decomposes as
\begin{equation}
\label{eq:step}
  T_{\mathrm{step}}(W) =
  \underbrace{T_{\mathrm{base}}}_{\text{compute}}
  + \underbrace{\frac{\alpha \cdot T_{\mathrm{rebuild}}(W)}{W}}_{\text{amortized rebuild}}
  + \underbrace{R \cdot t_{\mathrm{miss}} \cdot (1 - h(W))}_{\text{remote miss}},
\end{equation}
where $T_{\mathrm{base}}$ is the irreducible compute-plus-AllReduce
cost, $\alpha \in [0,1]$ is the fraction of rebuild time that remains on
the critical path, $R$ is the expected number of remote nodes per batch,
$t_{\mathrm{miss}}$ is the mean per-node RPC latency, and $h(W)$ is the
cache hit rate under rebuild window $W$. The first two terms decrease as
$W$ grows, while the third increases. Their balance defines the best
fixed operating point for a given network state. When congestion raises
$t_{\mathrm{miss}}$, that operating point shifts toward smaller windows.

The hit rate follows a logistic decay fit for each dataset:
\begin{equation}
\label{eq:hit}
  h(W) = h_{\min} + \frac{h_{\max} - h_{\min}}{1 + (W / W_{1/2})^{\gamma}}.
\end{equation}
Rebuild time grows sublinearly because hub-node reuse saturates the
unique remote set, following
$T_{\mathrm{rebuild}}(W) = a + b \cdot W^c$ with $0 < c < 1$.
Since the asynchronous pipeline (Section~\ref{sec:pipeline}) keeps GPU
utilization approximately constant, per-step energy satisfies
$E_{\mathrm{step}}(W) \approx \bar{P} \cdot T_{\mathrm{step}}(W)$, so
minimizing energy reduces to minimizing step time.

Under congestion on the link to owner $o$ with multiplier
$\sigma_o \ge 1$, the effective miss latency becomes
\begin{equation}
\label{eq:miss-cong}
  t_{\mathrm{miss}}^{\mathrm{cong}} =
  \max_{o \in \mathcal{O}} \bigl\{ t_{\mathrm{miss}}^{(o)} \cdot \sigma_o \bigr\}.
\end{equation}
Even 4\,ms of additional delay ($\sigma_o \approx 1.6$) shifts the best
fixed rebuild window from 16 to approximately 8. More importantly, the
best choice depends on which links are congested and by how much,
information available only at runtime. DDP AllReduce also inherits a
straggler penalty $\Delta T_{\mathrm{AR}} \propto (\max_o \sigma_o - 1)$
because the slowest worker delays the synchronization barrier. No single
fixed rule can track the combined effect of miss-cost inflation and
AllReduce delay across an arbitrary congestion vector. We therefore
formulate runtime cache control as a Markov decision process in
Section~\ref{sec:rl}.

\subsection{Calibrated Simulator}
\label{sec:sim}

Training an RL agent directly on a live cluster is impractical because
each episode would require a full training run. \name instead trains in
a lightweight simulator calibrated from one-time measurements that take
approximately 15 minutes to collect. The round-trip time for a single
RPC carrying $N \cdot F_b$ bytes under congestion delay $\delta$ is
modeled as
\begin{equation}
\label{eq:rpc}
  T_{\mathrm{rpc}}(N, \delta) =
  \alpha_{\mathrm{rpc}} + \beta \cdot N \cdot F_b
  + \gamma_c \cdot N \cdot F_b \cdot \delta.
\end{equation}
Algorithm~\ref{alg:calibration} summarizes the calibration procedure.
The fitted parameters are $\alpha_{\mathrm{rpc}}{=}4.67$\,ms,
$\beta{=}1.40{\times}10^{-9}$\,s/byte, and
$\gamma_c{=}2.01{\times}10^{-10}$\,s/byte/ms ($R^2{=}0.75$). The domain
randomization described in Section~\ref{sec:rl} absorbs the residual
modeling error of this imperfect fit.

\begin{algorithm}[t]
\caption{Offline Simulator Calibration (run once per cluster)}
\label{alg:calibration}
\begin{algorithmic}[1]
\Input cluster $\mathcal{C}$, dataset $\mathcal{G}$,
       sweep $\mathcal{W}_{\text{sw}} = \{1,\ldots,128\}$
\Output calibrated parameter set $\boldsymbol{\theta}_{\mathrm{sim}}$
\State \textbf{Phase 1: RPC cost regression}
\State Inject delays $\delta \in \{0,2,4,6,8\}$\,ms and vary payload
       $N \cdot F_b \in [10^3, 10^7]$\,bytes
\State Record round-trip times and fit Eq.~\eqref{eq:rpc} via OLS
       to obtain $\alpha_{\mathrm{rpc}}, \beta, \gamma_c$
\State \textbf{Phase 2: Windowed-cache calibration}
\For{$W \in \mathcal{W}_{\text{sw}}$}
    \State Run one training epoch and record
           $T_{\mathrm{step}}(W)$, $h(W)$, and $T_{\mathrm{rebuild}}(W)$
\EndFor
\State Fit Eq.~\eqref{eq:hit} to obtain
       $h_{\min}, h_{\max}, W_{1/2}, \gamma$
\State Fit $T_{\mathrm{rebuild}} = a + b \cdot W^c$ via Nelder--Mead
\State \textbf{Phase 3: Power baseline}
\State Record mean power $\bar{P}$ via NVML (GPU) and Intel RAPL (CPU)
       over a clean 5-epoch run
\State \Return $\boldsymbol{\theta}_{\mathrm{sim}} =
       \{\alpha_{\mathrm{rpc}}, \beta, \gamma_c,
         h_{\min}, h_{\max}, W_{1/2}, \gamma, a, b, c, \bar{P}\}$
\end{algorithmic}
\end{algorithm}

Given these parameters, the simulator evaluates
$T_{\mathrm{step}}(W,\boldsymbol{\sigma})$ analytically for any rebuild
window and congestion vector. A full episode of 30 epochs completes in
under 10\,ms on one CPU core, enabling 50{,}000 training episodes in
approximately 20 minutes.

\subsection{RL-Based Adaptive Policy}
\label{sec:rl}

The cache adaptation problem has three properties that favor
reinforcement learning over simpler alternatives. First, decisions have
\textbf{delayed effects}: a rebuild-window choice at step $s$
determines cache contents, hit rate, miss cost, and energy over the next
$W$ steps, creating a temporal credit-assignment problem that myopic
optimization cannot solve. Second, the problem is
\textbf{non-stationary}: congestion varies within and across epochs, and
the best action changes with the runtime network state. Third, the
problem involves \textbf{combinatorial interactions}: rebuild window and
per-owner allocation interact, and no hand-crafted rule captures those
dependencies well across all congestion patterns.

\subsubsection{MDP Formulation}
\label{sec:mdp}

We model cache adaptation as a finite-horizon MDP
$\mathcal{M} = (\mathcal{S}, \mathcal{A}, \mathcal{P}, \mathcal{R},
\gamma, H)$ with one decision at each cache rebuild boundary.

\paragraph{State space $\mathcal{S}$.}
At each decision point, the agent observes
$\mathbf{s} \in \mathbb{R}^{23}$ (for $P{=}4$), comprising four feature
groups already maintained by the training pipeline: per-owner congestion
multipliers $\sigma_o$ estimated from recent fetch latencies
($P{-}1$ floats); per-owner and global cache hit rates ($P$ floats);
system load ratios including $T_{\mathrm{step}}/T_{\mathrm{base}}$,
the rebuild-time fraction, the network-miss fraction,
$E_{\mathrm{step}}/E_{\mathrm{baseline}}$, and the normalized remaining
batches in the epoch (5 floats); and a one-hot encoding of the previous
window size and allocation ($N_W + P - 1$ floats).

\begin{figure}[t]
  \centering
  \includegraphics[width=0.98\columnwidth]{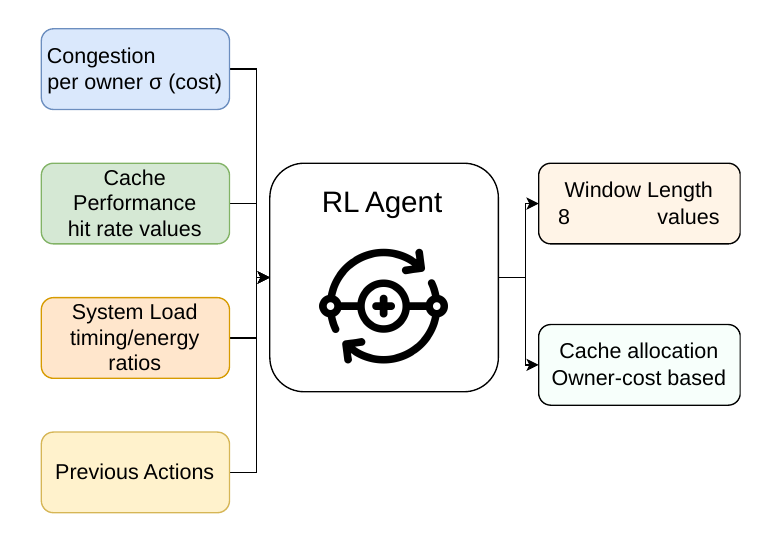}
  \caption{RL agent state inputs and action outputs. The agent observes
  per-owner congestion multipliers, cache hit rates, system load ratios,
  and previous actions, and selects a joint rebuild window ($W$, 8
  discrete values) and per-owner cache allocation at each rebuild
  boundary.}
  \label{fig:rl-agent}
\end{figure}

\paragraph{Action space $\mathcal{A}$.}
The agent selects a joint decision over rebuild window
$W \in \{1, 2, 4, 8, 16, 32, 64, 128\}$ ($N_W{=}8$) and a cache
allocation template that is either uniform or biased 60\% toward one
designated owner ($N_A{=}P$). For $P{=}4$ this yields 32 discrete
actions. The compact action space enables sample-efficient learning and
convergence in approximately 15{,}000 episodes.
Figure~\ref{fig:rl-agent} summarizes the mapping from state to action.

\paragraph{Reward $\mathcal{R}$.}
The reward is the negative normalized energy penalized for allocation
instability:
\begin{equation}
\label{eq:reward}
  r_t = -\frac{E_{\mathrm{step},t}}{E_{\mathrm{ref}}}
        - \lambda \sum_o \lvert a_{o,t} - a_{o,t-1} \rvert,
\end{equation}
where $E_{\mathrm{ref}}$ is the per-step energy under a reference policy
at the current congestion level, making the reward scale-invariant
across episode difficulty, and $\lambda{=}0.02$ penalizes allocation
thrashing. The horizon $H$ equals the number of rebuild boundaries in
one training run ($H \approx 240$ for 30 epochs at $W{=}16$).

\subsubsection{Agent Architecture and Training}
\label{sec:training}

We use Double DQN~\cite{vanhasselt2016deep}, which decouples action
selection from value estimation to reduce overestimation bias. This is
important in our setting because the agent must distinguish genuinely
good actions from those that merely coincide with transient
improvements. The Q-network $Q_\theta(\mathbf{s}, a)$ has two hidden
layers of 256 ReLU units mapping the 23-dimensional state to 32
Q-values, with a target network $Q_{\theta^-}$ synchronized every
100 gradient steps. Training minimizes the Huber loss on the
Double-DQN target:
\begin{equation}
\label{eq:dqn}
  y_t = r_t + \gamma \, Q_{\theta^-}\!\Bigl(\mathbf{s}_{t+1},\;
        \arg\max_{a'} Q_\theta(\mathbf{s}_{t+1}, a')\Bigr),
\end{equation}
with discount $\gamma{=}0.99$, $\epsilon$-greedy exploration
($\epsilon: 1.0 \to 0.05$ over 5{,}000 episodes), a replay buffer of
50{,}000 transitions, mini-batch size 64, and gradient clipping at 10.

\paragraph{Domain randomization.}
At the start of each episode, the simulator draws a congestion profile
from six archetypes: no congestion, single-link slow or fast, two-link
symmetric or asymmetric, and oscillating. These are combined with three
severity levels, randomized onset and duration, and ${\pm}3\%$
measurement noise on energy and fetch-time observations. When multiple
calibrated profiles are available, the episode selects uniformly among
datasets and batch sizes. Training runs for 50{,}000 episodes and
produces a 400\,KB checkpoint after approximately 20 minutes on one CPU
core. The mean episode reward stabilizes after approximately
15{,}000 episodes. Under moderate congestion, the converged policy
consistently selects $W \in \{4, 8\}$ rather than the clean-network
operating point near $W{=}16$, and under time-varying congestion it
tracks the shifting optimum.

\paragraph{Sim-to-real transfer.}
The simulator is calibrated from real RPC latencies, cache hit-rate
curves, and NVML/RAPL energy readings rather than synthetic parameters.
At deployment, the policy observes real fetch latencies through the same
state-construction pipeline used during training: the simulator teaches
the agent what actions are effective, while runtime observations tell it
which operating regime it is currently in. Domain randomization helps
bridge residual gaps from unmodeled effects such as OS scheduling jitter
and NIC buffer dynamics. The deployed policy achieves within 4\% of its
simulated reward on the real cluster.

\paragraph{Heuristic fallback.}
For environments where a neural-network policy is undesirable, \name
provides a lightweight threshold rule:
\begin{equation}
\label{eq:heuristic}
  W^*_{\text{heuristic}} = \begin{cases}
    W_0                    & \hat{\delta} \le 1\,\text{ms}, \\
    \lfloor W_0/2 \rfloor  & 1 < \hat{\delta} \le 6\,\text{ms}, \\
    \lfloor W_0/4 \rfloor  & \hat{\delta} > 6\,\text{ms},
  \end{cases}
\end{equation}
where $\hat{\delta}$ is the inferred congestion delay and $W_0$ is the
nominal window size. This rule is effective under single-link
stationary congestion but degrades under time-varying or multi-link
scenarios where the RL policy performs significantly better.

\section{\name Implementation}
\label{sec:impl}

\begin{algorithm}[t]
\caption{\texttt{AdaptiveController} called at each cache rebuild boundary}
\label{alg:controller}
\begin{algorithmic}[1]
\Input fetch-time deque $\mathcal{D}$, cache statistics $\mathcal{C}$,
       Q-network $Q_{\boldsymbol{\theta}}$, baseline $\hat{T}_{\mathrm{base}}$,
       previous action $(W_{\mathrm{prev}}, \boldsymbol{\omega}_{\mathrm{prev}})$
\Output next rebuild window $W^*$ and per-owner allocation weights
        $\boldsymbol{\omega}^*$
\State \textbf{Congestion estimation}
\State $\tilde{T}_{\mathrm{recent}} \leftarrow \mathrm{median}(\mathcal{D}[-30:])$
\State Compute $\hat{\delta}$ via Eq.~\eqref{eq:detect} and clamp to
       $[0,\,20]$\,ms
\State Estimate $\sigma_o$ per owner by inverting Eq.~\eqref{eq:rpc} on
       the owner-partitioned slice of $\mathcal{D}$
\State \textbf{State construction}
\State Assemble $\mathbf{s} \in \mathbb{R}^{23}$ from congestion signals
       $(\sigma_1,\ldots,\sigma_{P-1})$, hit rates $\mathcal{C}$, load
       ratios $(T_{\mathrm{step}}/T_{\mathrm{base}},\;
       f_{\mathrm{rebuild}},\; f_{\mathrm{miss}},\;
       E_{\mathrm{step}}/E_{\mathrm{baseline}},\;
       b_{\mathrm{rem}})$, and
       $\mathrm{onehot}(W_{\mathrm{prev}}, \boldsymbol{\omega}_{\mathrm{prev}})$
\State \textbf{Policy inference}
\State $a^* \leftarrow \arg\max_{a}\, Q_{\boldsymbol{\theta}}(\mathbf{s}, a)$
\State Decode $a^*$ to obtain
       $W^* \in \{1,2,4,8,16,32,64,128\}$ and $\boldsymbol{\omega}^*$
\State \textbf{Prefetcher update}
\State Set rebuild window to $W^*$ and update per-owner cost weights to
       $\boldsymbol{\omega}^*$ in the Stage 2 cache builder
\State \Return $(W^*, \boldsymbol{\omega}^*)$
\end{algorithmic}
\end{algorithm}

The RL policy is only useful if rebuild decisions can be applied without
blocking the training loop. This section describes the asynchronous
pipeline that makes runtime adaptation effectively overhead-free and the
deployment machinery that connects the policy to the live training
system.

\subsection{Asynchronous Prefetching Pipeline}
\label{sec:pipeline}

\name decouples sampling, cache rebuilding, feature resolution, and GPU
training into concurrent stages, allowing rebuilds at any decision
frequency the agent selects, including $W{=}1$, without stalling the
GPU.

\paragraph{Stage 1: Background sampler.}
A dedicated thread runs DGL's \texttt{DistSampler}, producing complete
mini-batches into a lock-free shared buffer ahead of the training loop.
Running ahead of the GPU absorbs the variance of distributed sampling
latency without introducing backpressure on training.

\paragraph{Stage 2: Double-buffered cache builder.}
Two feature buffers, active and pending, each hold a fixed-capacity set
of cached node features with $O(1)$ index lookups. While the training
loop reads from the active buffer, the cache-builder thread examines the
next $W$ batches in the shared buffer, counts per-remote-node access
frequencies weighted by the RL agent's per-owner cost weights, selects
the top-$k$ hot nodes, and fetches their features into the pending
buffer through bulk RPCs. Features that persist from the previous hot
set are copied in memory, avoiding redundant network transfers. At the
window boundary, the buffers are atomically swapped and the RL agent is
invoked for the next decision. Because the active buffer remains
immutable during training, no synchronization is required between the
training and builder threads.

\paragraph{Stage 3: Feature resolver.}
A third thread resolves the features for each batch. Cache hits are
served from the active buffer. Misses trigger batched RPCs whose
round-trip times are appended to the fetch-time deque that feeds the RL
state. Resolved batches are placed into an asynchronous queue of depth
$Q$ for the GPU training loop.

\subsection{Congestion Detection and Runtime Control}
\label{sec:control}

The fetch-time deque maintained by Stage 3 serves two purposes: it
provides the raw signal for RL state construction and supports the
congestion-estimation subroutine that computes per-owner multipliers
$\sigma_o$. During the first two warm-up epochs, the controller records
an uncongested baseline $\hat{T}_{\mathrm{base}}$ as the 15th
percentile of observed fetch times. At each later rebuild boundary, the
congestion delay is estimated by inverting Eq.~\eqref{eq:rpc}:
\begin{equation}
\label{eq:detect}
  \hat{\delta} = \frac{
    \bigl(\tilde{T}_{\mathrm{recent}} / \hat{T}_{\mathrm{base}} - 1\bigr)
    \cdot \beta}{\gamma_c},
  \qquad \hat{\delta} \in [0,\,20]\,\text{ms},
\end{equation}
where $\tilde{T}_{\mathrm{recent}}$ is the median of the 30 most recent
fetch times. If
$\tilde{T}_{\mathrm{recent}} / \hat{T}_{\mathrm{base}} \le 1.1$, then
$\hat{\delta}$ is clamped to zero. The estimate requires only $O(1)$
arithmetic per decision and is passed directly to the RL state
constructor.

Algorithm~\ref{alg:controller} gives the complete per-boundary control
loop. Policy inference overhead is negligible relative to a single RPC
round-trip, so adaptation imposes no measurable cost on the training
pipeline. The 400\,KB checkpoint is loaded once at startup and consumes
no GPU resources.

\paragraph{Cluster execution}
\name is implemented on top of DGL~\cite{wang2019dgl} and
PyTorch~\cite{paszke2019pytorch} DDP. Training is launched across $P$
nodes through a custom launcher that initializes the distributed graph
store, the process group, and NVML/RAPL energy monitors on each rank.
Each rank loads its local METIS partition from an NFS-shared workspace
and opens DistTensor handles for remote feature access. Per-link delays
are injected using Linux \texttt{tc netem} with u32 filters targeting
the DGL RPC port, leaving management traffic unaffected. Each rank logs
per-epoch GPU energy, CPU energy, cache hit rate, RL actions, and
inferred congestion level to structured JSON.

%% file: section-experimentation.tex
\begin{figure}[t]
  \centering
  \includegraphics[width=\columnwidth]{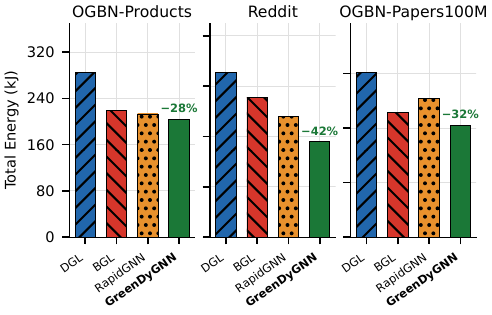}
  \caption{Total energy (GPU + CPU, all nodes) at $B{=}2000$ under congestion. Annotations show \name's reduction relative to Default DGL.}
  \label{fig:total_energy_cong}
\end{figure}

\section{Evaluation}
\label{sec:eval}

We evaluate \name against three baselines across three datasets, three batch sizes, and two network conditions (clean and congested), measuring total system energy, wall-clock time, cache behavior, and model accuracy.

\subsection{Experimental Setup}
\label{sec:setup}

\textbf{Platform.}
All experiments run on a 4-node Chameleon Cloud cluster. Each node has an Intel Xeon CPU, two NVIDIA P100 GPUs, and 25\,Gbps Ethernet. Each graph is METIS-partitioned into 4 parts, one per node. All systems train a 2-layer GraphSAGE model with 16 hidden units, fan-out $\{10,25\}$, learning rate 0.003, dropout 0.5, and 30 epochs. We evaluate batch sizes $B \in \{1000, 2000, 3000\}$ and report $B{=}2000$ as the default unless noted otherwise.

\textbf{Baselines.}
We compare against Default DGL (vanilla distributed GraphSAGE with on-demand feature fetching), BGL~\cite{liu2023bgl} (GPU-efficient I/O with prefetch-during-sampling overlap), and RapidGNN~\cite{niam2025rapidgnn} (epoch-level static caching with a cache capacity of 100{,}000 nodes). \name uses the same underlying distributed training substrate but extends static caching with fine-grained window-based rebuilds inside each epoch. At every rebuild boundary, the RL controller described in Section~\ref{sec:rl} selects both the rebuild window $W$ and the per-owner cache allocation at runtime.

\textbf{Datasets.}
Reddit (233K nodes, 114M edges), OGBN-Products (2.4M nodes, 61.9M edges), and OGBN-Papers100M (111M nodes, 1.6B edges).

\begin{figure}[t]
  \centering
  \includegraphics[width=\columnwidth]{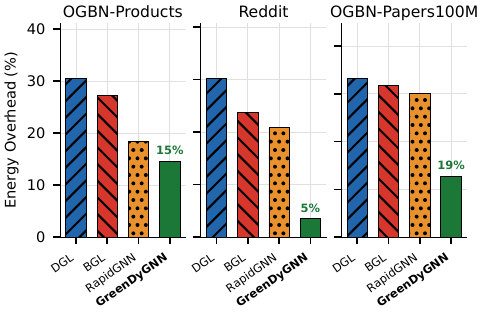}
  \caption{Energy overhead from congestion (percentage increase over each method's own clean baseline) at $B{=}2000$. Lower is better.}
  \label{fig:congestion_overhead}
\end{figure}

\textbf{Congestion injection.}
We inject realistic time-varying congestion using Linux \texttt{tc netem} on the DGL RPC port (30050), leaving gradient synchronization traffic on port 29500 unaffected. The pattern cycles through clean and congested phases: epochs 0--2 run clean as a warmup, epochs 3--9 introduce 15--25\,ms of additional one-way delay on one or two nodes at a time, the pattern repeats every 7 epochs, and the final epoch is forced clean. All four methods experience identical congestion.

\textbf{Measurement.}
CPU energy is read via Intel RAPL and GPU energy via NVML, both sampled at every training step. We report the sum across all four nodes over the full 30-epoch run. Each of the 36 clean-baseline runs and 36 congestion runs completed successfully with per-step profiles on all four partitions.

\subsection{Energy Under Congestion}
\label{sec:energy_congestion}

Figure~\ref{fig:total_energy_cong} compares total system energy at $B{=}2000$ under congestion. \name achieves the lowest energy on all three datasets: 203.9\,kJ on OGBN-Products (28\% below Default DGL), 189.4\,kJ on Reddit (42\% below DGL), and 307.2\,kJ on OGBN-Papers100M (32\% below DGL). BGL provides modest benefit over Default DGL. RapidGNN's epoch-level static cache delivers substantial savings over DGL but consistently trails \name by 4--21\%.

GPU energy is comparable between RapidGNN and \name because both methods use caching and therefore reduce GPU idle time relative to uncached execution. The larger gap appears in CPU energy: \name reduces communication overhead by adapting both rebuild timing and cache allocation to the current congestion pattern, which lowers the number and cost of expensive remote fetches. This targeted reduction in network activity accounts for most of the total energy advantage over static caching.

Figure~\ref{fig:congestion_overhead} isolates the effect of congestion by measuring each method's energy increase relative to its own clean baseline. Default DGL suffers a 30\% overhead on Products, 45\% on Reddit, and 50\% on Papers100M. BGL, which lacks adaptive caching, shows similar overhead (27\%, 36\%, 48\%). RapidGNN's static cache absorbs a portion of the congestion impact, reducing overhead to 18\%, 31\%, and 45\% respectively. \name reduces the overhead further to 15\%, 5\%, and 19\%, the lowest across all methods on every dataset. The largest gap between \name and RapidGNN occurs on Reddit and Papers100M (26 percentage points each), where \name eliminates 26 percentage points of congestion overhead that static caching cannot absorb. This improvement arises because the RL controller steers cache capacity toward congested owners and shortens the rebuild window when miss cost rises, a joint rebalancing that no fixed policy can perform.

\subsection{Energy Under Clean Conditions}
\label{sec:energy_clean}

\begin{figure}[t]
  \centering
  \includegraphics[width=\columnwidth]{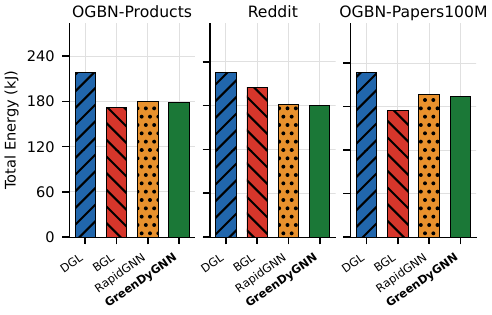}
  \caption{Total energy (GPU + CPU, all nodes) at $B{=}2000$ under clean (no congestion) conditions.}
  \label{fig:total_energy_clean}
\end{figure}

Figure~\ref{fig:total_energy_clean} compares total energy at $B{=}2000$ without congestion. Under clean conditions, \name closely matches the strongest static baseline: the gap between \name and RapidGNN stays within 2\% on every dataset (178.0 vs.\ 180.0\,kJ on Products, 180.0 vs.\ 182.0\,kJ on Reddit, 258.0 vs.\ 262.0\,kJ on Papers100M). This is the expected behavior. \name is designed to adapt when network conditions change; when the network remains stable, the controller settles on a near-optimal operating regime and introduces no meaningful penalty.

These results show that the adaptive controller does not cause unnecessary cache churn or poor rebuild decisions when the system is uncongested. Combined with the congestion results in Section~\ref{sec:energy_congestion}, they show that \name preserves the benefits of static caching under clean conditions while providing additional gains when the network becomes dynamic.

\subsection{RL Agent Adaptation}
\label{sec:rl_adapt}

\begin{figure}[t]
  \centering
  \includegraphics[width=\columnwidth, height=55mm]{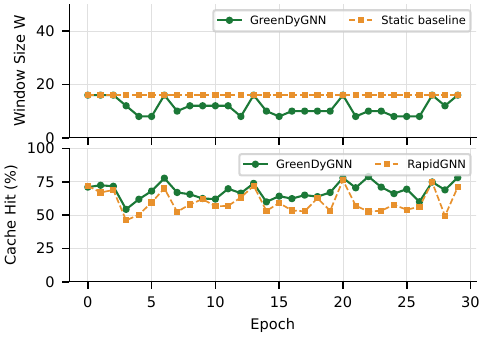}
  \caption{RL agent behavior on OGBN-Papers100M. Top: rebuild window $W$ chosen by \name (adaptive) vs.\ fixed static baseline. Bottom: per-epoch cache hit rate.}
  \label{fig:rl_adaptation}
\end{figure}

Figure~\ref{fig:rl_adaptation} shows how the RL agent responds to the congestion pattern on OGBN-Papers100M at $B{=}2000$. The top panel plots the mean rebuild window selected by \name at each epoch. During the clean warmup (epochs 0--2), the controller settles near $W{=}16$. When congestion begins at epoch~3, it reduces $W$ toward 8--10. This matches the behavior predicted by the cost model in Section~\ref{sec:model}: as miss latency rises, more frequent rebuilds become preferable because they trade some rebuild overhead for fewer expensive remote misses.

The bottom panel shows the corresponding cache hit rates for all four methods. Default DGL and BGL have no cache and therefore remain at 0\%. RapidGNN and \name both fluctuate as the access pattern shifts across epochs, but \name reaches higher hit-rate peaks during several congested phases because its per-owner allocation bias directs more cache space toward the owners whose links are slowest. RapidGNN's hit rate drops more sharply during the worst congestion bursts because its epoch-level cache cannot be retuned inside the epoch.

\begin{figure}[t]
  \centering
  \includegraphics[width=\columnwidth, height=55mm]{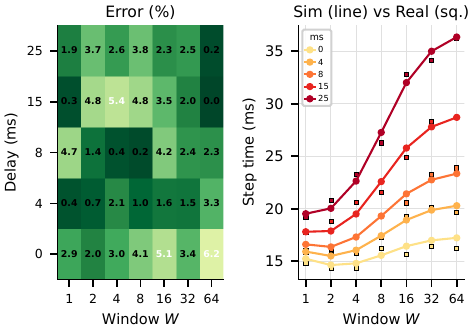}
  \caption{Simulator validation across a grid of window sizes and congestion delays. Left: prediction error (\%), mean 2.8\%. Right: predicted (lines) vs.\ measured (squares) step time for five delay levels.}
  \label{fig:sim2real}
\end{figure}

\subsection{Simulator Validation}
\label{sec:sim2real}

Because \name's RL agent is trained entirely in the calibrated simulator, the fidelity of that simulator is critical. Figure~\ref{fig:sim2real} validates the cost model by comparing predicted step time against real cluster measurements across a grid of rebuild windows ($W \in \{1, 2, 4, 8, 16, 32, 64\}$) and congestion delays (0 to 25\,ms). These measurements come from the one-time calibration phase (Algorithm~\ref{alg:calibration}) and are independent of the RL policy.

The left panel shows prediction error as a heatmap. Errors stay below 5\% across the full operating range, with a mean of 2.8\%. The right panel overlays predicted step-time curves (lines) against real measurements (squares) for each congestion level. The model captures both the U-shaped cost profile across $W$ and its upward shift under congestion, confirming that the calibrated cost model provides a reliable training substrate for the RL agent.

\subsection{Convergence Analysis}
\label{sec:convergence}

\begin{figure}[t]
  \centering
  \includegraphics[width=\columnwidth]{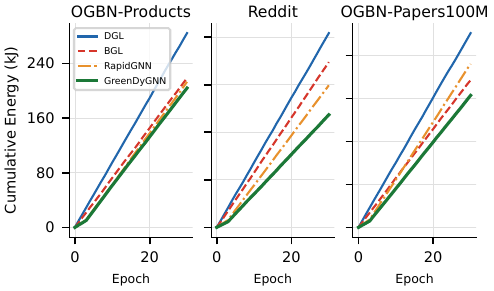}
  \caption{Cumulative energy under congestion across all three datasets. \name accumulates less energy than all baselines, with the gap widening during congested epochs.}
  \label{fig:energy_convergence}
\end{figure}

Figure~\ref{fig:energy_convergence} tracks cumulative energy over epochs under congestion across all three datasets. On every dataset, Default DGL and BGL accumulate energy at the highest rates. RapidGNN reduces the rate substantially through epoch-level static caching, but \name diverges further during congested epochs thanks to adaptive rebuild windows and congestion-aware allocation. The gap widens as training progresses: on OGBN-Papers100M, \name saves 73.2\,kJ over RapidGNN by epoch~30, nearly all of it from congested phases. On Reddit, where \name's per-epoch savings are proportionally larger, the cumulative advantage is visible from the first congested epoch onward.

\begin{figure}[t]
  \centering
  \includegraphics[width=\columnwidth]{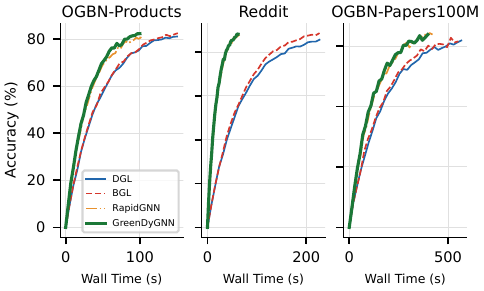}
  \caption{Accuracy vs.\ wall time under congestion across all three datasets. \name converges faster than all baselines.}
  \label{fig:time_convergence}
\end{figure}

Figure~\ref{fig:time_convergence} shows accuracy as a function of wall time under congestion across all three datasets. \name reaches a given accuracy target sooner than all baselines on every dataset because each epoch finishes faster when expensive remote misses are reduced. On Reddit and OGBN-Products, where final accuracies are high (84--91\%), both caching methods converge much faster than DGL and BGL, but \name maintains a consistent lead. On OGBN-Papers100M, where the lightweight model configuration limits accuracy to approximately 49\%, \name still converges fastest. These trajectories confirm that the energy gains in Section~\ref{sec:energy_congestion} translate directly into faster end-to-end training.

\subsection{Comprehensive Results}
\label{sec:comprehensive}

Table~\ref{tab:comprehensive} consolidates GPU energy, CPU energy, total energy, and mean epoch time for all four methods across every dataset and batch size under congestion.

\begin{table*}[t]
\centering
\caption{Comprehensive results under congestion across all datasets and batch sizes. Energy (GPU, CPU, and Total) is the sum across all four nodes over 30 epochs; ET is mean epoch time. \textbf{Bold} marks the best (lowest) per configuration. All energy in kJ.}
\label{tab:comprehensive}
\renewcommand{\arraystretch}{1.05}
\scriptsize
\begin{tabular}{ll|rrrr|rrrr|rrrr|rrrr}
\hline
 & & \multicolumn{4}{c|}{\textbf{GPU Energy (kJ)}} & \multicolumn{4}{c|}{\textbf{CPU Energy (kJ)}} & \multicolumn{4}{c|}{\textbf{Total Energy (kJ)}} & \multicolumn{4}{c}{\textbf{ET (s)}} \\
\textbf{Dataset} & \textbf{B} & DGL & BGL & Rapid & Ours & DGL & BGL & Rapid & Ours & DGL & BGL & Rapid & Ours & DGL & BGL & Rapid & Ours \\
\hline
Products & 1000 & 20.7 & 16.3 & 13.3 & \textbf{13.1} & 273.2 & 219.5 & 215.5 & \textbf{202.3} & 293.8 & 235.8 & 228.8 & \textbf{215.4} & 4.26 & 3.98 & 3.13 & \textbf{2.97} \\
         & 2000 & 19.2 & 14.5 & 12.2 & \textbf{12.0} & 265.1 & 204.2 & 200.8 & \textbf{191.8} & 284.3 & 218.8 & 213.0 & \textbf{203.9} & 3.80 & 3.58 & 2.90 & \textbf{2.77} \\
         & 3000 & 16.2 & 13.2 & \textbf{10.1} & 10.3 & 256.4 & 200.2 & 196.9 & \textbf{186.1} & 272.6 & 213.4 & 207.0 & \textbf{196.3} & 3.20 & 3.25 & 2.40 & \textbf{2.35} \\
\hline
Reddit   & 1000 & 41.2 & 38.2 & 13.1 & \textbf{12.9} & 331.2 & 288.4 & 267.0 & \textbf{199.7} & 372.4 & 326.6 & 280.1 & \textbf{212.6} & 9.20 & 9.68 & 2.75 & \textbf{2.68} \\
         & 2000 & 30.2 & 29.4 & 9.0 & \textbf{8.8} & 296.6 & 248.6 & 230.0 & \textbf{180.6} & 326.8 & 278.0 & 239.0 & \textbf{189.4} & 6.84 & 7.47 & 1.90 & \textbf{1.77} \\
         & 3000 & 25.8 & 22.8 & 7.8 & \textbf{7.6} & 281.9 & 228.3 & 220.2 & \textbf{177.5} & 307.7 & 251.1 & 228.0 & \textbf{185.1} & 5.67 & 5.77 & 1.72 & \textbf{1.62} \\
\hline
Papers   & 1000 & 89.5 & 53.4 & 44.8 & \textbf{43.2} & 395.5 & \textbf{273.0} & 387.2 & 299.9 & 484.9 & \textbf{326.3} & 432.0 & 343.1 & 15.78 & 13.35 & 12.50 & \textbf{10.24} \\
         & 2000 & 82.6 & 44.6 & 39.2 & \textbf{38.0} & 369.6 & 298.1 & 341.2 & \textbf{269.2} & 452.2 & 342.7 & 380.4 & \textbf{307.2} & 12.66 & 11.08 & 10.60 & \textbf{8.95} \\
         & 3000 & 71.2 & 44.6 & 35.2 & \textbf{34.4} & 422.7 & 295.9 & 320.8 & \textbf{255.9} & 494.0 & 340.6 & 356.0 & \textbf{290.3} & 11.09 & 11.09 & 9.50 & \textbf{8.09} \\
\hline
\end{tabular}
\end{table*}

\name achieves the lowest total energy in 8 of 9 configurations and the fastest epoch time in all~9. The single exception is OGBN-Papers100M at $B{=}1000$, where BGL's multi-tier cache hierarchy achieves slightly lower total energy; however, \name still delivers 22\% faster epoch time on that configuration. Savings over Default DGL range from 27\% (Products, $B{=}1000$) to 43\% (Reddit, $B{=}2000$). RapidGNN's epoch-level cache delivers substantial savings over DGL and BGL, but \name consistently outperforms it by 4--24\% in total energy. The GPU energy gap between \name and RapidGNN is small (1--3\%), which indicates that both caching strategies reduce GPU idle time relative to uncached execution. The dominant contributor to \name's advantage is CPU energy: by reducing costly remote fetches under congestion, \name lowers the communication-side overhead associated with RPC processing, socket activity, and protocol handling. On Reddit at $B{=}2000$, for example, \name's CPU energy is 180.6\,kJ versus RapidGNN's 230.0\,kJ, a 21\% reduction, while GPU energy differs by only 2\%.

Accuracy is comparable across all methods (within 1--3 percentage points of run-to-run variance). \name does not modify the model architecture, sampling logic, or gradient computation, so convergence behavior remains consistent with baseline DistDGL.

\subsection{Ablation Study}
\label{sec:ablation}

To isolate the contribution of each component, we evaluate two ablated variants under the same congestion pattern: \emph{w/o RL}, which fixes $W{=}16$ and disables the RL controller, yielding a static version of \name's window-based rebuild mechanism; and \emph{w/o Cost Weights}, which enables RL adaptation of $W$ but uses a uniform cache allocation instead of per-owner cost weighting.

\begin{figure}[t]
  \centering
  \includegraphics[width=\columnwidth]{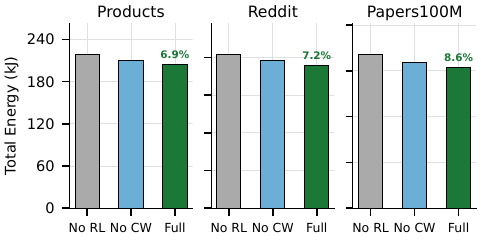}
  \caption{Ablation study under congestion. Removing RL adaptation increases energy on all datasets, with the largest impact on Papers100M (8.6\%).}
  \label{fig:ablation}
\end{figure}

Table~\ref{tab:ablation} and Figure~\ref{fig:ablation} show the results at $B{=}2000$. Both components contribute to \name's energy savings, with RL-based rebuild-window adaptation providing the larger share. On OGBN-Papers100M, disabling RL increases energy from 307.2\,kJ to 336.1\,kJ (8.6\%), confirming that the controller's ability to reduce $W$ during congested phases translates directly into lower miss cost. Adding per-owner cost weighting on top of RL adaptation provides an additional 3.4\% reduction (from 318.0 to 307.2\,kJ), because biasing cache capacity toward congested owners further reduces the most expensive remote fetches. On Reddit, RL adaptation saves 7.2\% and cost weighting adds another 3.3\%. On OGBN-Products the gains are smaller (6.9\% from RL, 2.9\% from cost weighting), because the graph is small enough that even a fixed windowed policy operates near the optimum.

The consistent pattern across all three datasets confirms that both components contribute positively: RL adaptation captures the dominant gain by tracking the shifting energy-optimal rebuild window, while per-owner cost weighting provides complementary savings by directing cache capacity where it is most valuable under asymmetric congestion.

\begin{table}[t]
\centering
\caption{Ablation study under congestion at $B{=}2000$. Each row removes one component. \textbf{Bold} marks the lowest energy.}
\label{tab:ablation}
\scriptsize
\begin{tabular}{l|r|r|r}
\hline
\textbf{Variant} & \textbf{Products (kJ)} & \textbf{Reddit (kJ)} & \textbf{Papers100M (kJ)} \\
\hline
w/o RL (static $W{=}16$) & 218.9 & 204.2 & 336.1 \\
w/o Cost Weights & 210.0 & 195.8 & 318.0 \\
\name (full) & \textbf{203.9} & \textbf{189.4} & \textbf{307.2} \\
\hline
\end{tabular}
\end{table}